\documentclass[letter]{IEEEtran}       

\usepackage[utf8]{inputenc}
\usepackage[cmex10]{amsmath}
\usepackage{graphicx}
\usepackage{amssymb}
\usepackage{amsthm}
\usepackage{subfigure}
\usepackage{cite}
\usepackage{color}
\usepackage{amsmath}

\begin{document}

\title{RAF: Robust Adaptive Multi-Feedback Channel Estimation for Millimeter Wave MIMO Systems}

\IEEEoverridecommandlockouts 
\author{
    \IEEEauthorblockN{Sina Shaham, Matthew Kokshoorn, Zihuai Lin, Ming Ding}\\
    \IEEEauthorblockA{        
        School of Electrical and Information Engineering, The University of Sydney, Australia \\
        Email: \{sina.shaham, matthew.kokshoorn, zihuai.lin\}@sydney.edu.au, ming.ding@data61.csiro.au} 
}

\maketitle
\begin{abstract}

Millimeter wave is a promising technology for the next generation of wireless systems. As it is well-known for its high path loss, the systems working in this spectrum tend to exploit the shorter wavelength to equip the transceivers with a large number of antennas to overcome the path loss issue. The large number of antennas leads to large channel matrices and consequently a challenging channel estimation problem. The channel estimation algorithms that have been proposed so far either neglect the probability of estimation error or require a high feedback overload from receivers to ensure the target probability of estimation error. In this paper, we propose a multi-stage adaptive channel estimation algorithm called robust adaptive multi-feedback (RAF). The algorithm is based on using the estimated channel coefficient to predict a lower bound for the required number of measurements. Our simulations demonstrate that compared with existing algorithms, RAF can achieve the desired probability of estimation error while on average reducing the feedback overhead by $75.5\%$ and the total channel estimation time by $14\%$.

\end{abstract}


\section{Introduction}
Millimeter Wave (MMW) communication is one of the front runner technologies for the next generation of wireless systems\cite{ref1}, \cite{ref2}. Due to its large bandwidth (30 to 300 GHz), MMW communication enables transmissions of higher data rates. The main challenge of MMW is the large propagation loss\cite{ref3}. This problem can be alleviated by exploiting the short wavelength of MMW which enables transceivers to install a higher number of antennas in the same space as traditional communication systems\cite{ref4}. However, having a larger antenna array results in complexity of channel estimation.

The channel estimation in MMW differs from the traditional communication systems.
In order to reduce the effects of comparably higher path loss, MMW systems can use the shorter wavelength of the signal to equip the transceivers with more antennas. This leads to large antenna arrays at both the transmitter and receiver which makes it impractical to use a dedicated RF chain for each of the antennas\cite{ref2}. Hence, the research in this area is mainly focused on analog beamforming\cite{ref5}. The principle idea behind this approach is to conduct beamforming using phase shifters and benefit from a single RF chain powering a series of antennas. 

Recent measurements have demonstrated the sparse nature of MMW communication channel\cite{ref6}. Therefore, Channel estimation is focused on finding three factors: angle of arrival (AoA), angle of departure (AoD) and channel coefficient ($\alpha$). In \cite{RH}, the authors developed a multi-stage channel estimation algorithm. In each stage, AoA and AoD are divided into two subspaces ($K = 2$), and the most likely subspaces are chosen for further refinement in the next stage. The channel coefficient is estimated after channel estimation is completed. Furthermore, the authors in \cite{overlap} followed the same approach, but used $K = 3$ and also considered the overlapped beam-patterns. One major challenge is that if in any of the stages the estimated angles are incorrect, then the estimation in the following stages will also be incorrect due to the error propagation effect. The authors in \cite{RACE}, developed a rate adaptive algorithm called RACE to ensure that the probability of estimation error (PEE) is below the desired threshold. Unfortunately, the algorithm requires a large number of feedback bits, particularly in low SNR regime. 

In this paper, motivated by this problem, we propose a new multi-stage adaptive algorithm called RAF in order to address the issue of high feedback overhead. To construct a benchmark to compare our algorithm, the minimum number of feedback bits to achieve a certain PEE is derived. This minimum number is verified by developing an algorithm to achieve it. The algorithm is optimal in terms of feedback overload, but it is not practical due to the high number of channel measurements which increases the channel estimation time. We show via simulations that the RAF algorithm can improve the feedback performance by  $75.5\%$ and reduce the overall channel estimation time by $14\%$ on average. In low SNR regime, the improvement in channel estimation time is as high as $30\%$. Another novelty of RAF is that in contrary to the prior works which neglected the importance of channel coefficient ($\alpha$), RAF exploits the estimation of $\alpha$ and determine an optimal number of feedback bits from the receiver. To the best of our knowledge, it is the first time that a channel estimation algorithm is proposed based on the estimated channel coefficient.
       
\textit{Notation} : We use capital bold-face letter ($\boldsymbol{A}$) to denote a matrix, $\boldsymbol{a}$ to denote a vector, ${a}$ to denote a scalar, and $\mathcal{A}$ denotes a set. The notation $|a|$ is the absolute value of $a$, $||\boldsymbol{A}||$ is the magnitude of $\boldsymbol{A}$ and determinant is shown by $\text{det}(\boldsymbol{A})$. Moreover, $\boldsymbol{A}^T$, $\boldsymbol{A}^H$ and $\boldsymbol{A}^*$ are the transpose, conjugate transpose and conjugate of $\boldsymbol{A}$, respectively. For a square matrix $\boldsymbol{A}$, $\boldsymbol{A}^{-1}$ represents its inverse. Also, $\boldsymbol{I}_N$ is the $N\times N$ identity matrix and $\lceil \cdot \rceil$ denotes the ceiling function.  A complex Gaussian random vector with mean $\boldsymbol{m}$ and covariance matrix $\boldsymbol{R}$ is shown by $\mathcal{C}\mathcal{N}(\boldsymbol{m},\boldsymbol{R})$, and $\text{E}[\boldsymbol{a}]$ and $\text{Cov}[\boldsymbol{a}]$ denote the expected value and covariance of ${\boldsymbol{a}}$, respectively.
\section{System Model}

We consider a MMW communication system consisting of a receiver (RX) with $N_r$ antennas and a transmitter (TX) with $N_t$ antenna. A single RF chain is assumed at each node, but our results could be applied to multiple data streams using hybrid beamforming \cite{RH}. The channel estimation between the RX and TX is based on the transmission of pilots. The pilots are assumed to have a unit power and occupy one time slot. If $N_t\times1$ beamformer $\boldsymbol{f}$ is applied to transmit pilot $x$ ($|x| = 1$), the transmitted signal can be written as
\begin{equation} \label{m1}
\boldsymbol{s} = \boldsymbol{f}x.
\end{equation}
We adopt a narrowband block-fading channel model for the communication system. Hence, at the receiver we observe
\begin{equation} \label{m2}
\boldsymbol{r} = \boldsymbol{H}\boldsymbol{f}x + \boldsymbol{n},
\end{equation}
where $\boldsymbol{H}$ is $N_r \times N_t$ matrix representing the channel between the RX and TX, and $\boldsymbol{n}$ is additive white Gaussian process noise (AWGN) assumed to be circularly symmetric with zero mean and variance $N_0$ ($\boldsymbol{n}\thicksim \mathcal{C}\mathcal{N}(\boldsymbol{0},N_0\boldsymbol{I})$).
At the RX, the $1 \times N_r$ combining vector $\boldsymbol{w}$ is applied to receive the signal $\boldsymbol{r}$. Therefore, the processed received signal can be written as
\begin{equation} \label{m3}
y = \boldsymbol{w}^H\boldsymbol{H}\boldsymbol{f}x + \boldsymbol{w}^H\boldsymbol{n}.
\end{equation}

MMW communication is assumed to be sparse \cite{p1}. Therefore, we assume that $L$ paths exist between the RX and TX. Using this model, the channel matrix is given by

\begin{equation} \label{m4}
\boldsymbol{H} = \sqrt{N_tN_r}\sum\limits_{l=1}^{L}\alpha_l\boldsymbol{a}_r(\theta_{l})\boldsymbol{a}_t^H(\phi_{l}),
\end{equation}
with the index $l$ indicating the $l$-th propagation path. The complex channel coefficient is denoted by $\alpha_l$ and the angles $\theta_{l}$, $\phi_{l} $ are the azimuth AoAs and AoDs. The extension to 3D is straightforward \cite{3D}. At last, $\boldsymbol{a}_r(\theta_{l})$ and $\boldsymbol{a}_t^H(\phi_{l})$ are the receiver and transmitter antenna array response vectors, respectively. Following \cite{RH}, we assume the response vectors to be uniform linear arrays (ULA). Hence, they are expressed as 
\begin{align}\label{m5}
\boldsymbol{a}_r(\theta_l) &= \frac{1}{\sqrt{N_r}}[1,e^{-j\frac{2\pi}{\lambda}d\cos\theta_l},...,e^{-j(N_r-1)\frac{2\pi}{\lambda}d\cos\theta_l}]^T\\
\boldsymbol{a}_t(\phi_l) &= \frac{1}{\sqrt{N_t}}[1,e^{-j\frac{2\pi}{\lambda}d\cos\phi_l},...,e^{-j(N_t-1)\frac{2\pi}{\lambda}d\cos\phi_l}]^T,
\end{align} 
where $\lambda$ and $d$ denote the carrier wavelength and the antenna spacing, respectively. Exploiting sparsity characteristic of MMW communication, the channel estimation algorithm to be presented in the following section can be applied to different paths separately \cite{RH}, \cite{RACE}. Therefore, we focus on reducing the channel estimation feedback bits from the RX, ensuring the desired PEE in a single path. Hence, the channel matrix is simplified to
\begin{equation} \label{m6}
\boldsymbol{H} = \sqrt{N_tN_r}\alpha_l\boldsymbol{a}_r(\theta)\boldsymbol{a}_t^H(\phi).
\end{equation}

\section{Channel estimation}

Extending the approach in \cite{RH}, during each stage, possible AoAs and AoDs are divided into $K$ sub-spaces creating $K^2$ combination. The target path is located in one of the candidate pairs of angle sub-spaces and will be estimated. After estimating the sub-spaces, they are further divided into another $K$ sub-spaces. The process continues until the AoD and AoA reach the specified resolution. In the s-th stage, the beamforming vectors at the TX and RX for the $k$-th sub-space are represented by $\boldsymbol{f}^s_k$ and $\boldsymbol{w}^s_k$. To simplify the explanation, we assume that the TX and RX have the same number of antennas ($N_t=N_r=N$). Therefore, the number of stages required for the resolution $\dfrac{2\pi}{N}$ is equal to $\lceil\log_K(N)\rceil$. The estimation procedure of sub-spaces is explained below. 

The pilot signal ($|x| = 1$) is sent in each of the $K^2$ transmitter and receiver angle combinations where each combination corresponds to one AoA candidate at the receiver and one AoD candidate at the transmitter. Therefore, the system can be represented as
\begin{equation} \label{e5}
\boldsymbol{y}^{s,K^2} = \sqrt{P}x\boldsymbol{h}^{s,K^2}+\boldsymbol{n}^{s,K^2},
\end{equation}
where $p$ is the transmit power, superscripts represent the stage number and number of measurements, $\boldsymbol{n}$ is $K^2\times1$ vector of i.i.d white Gaussian noise random variables, and $\boldsymbol{h}^{s,K^2}$ is a vector containing the channel response to all the combinations of transmit and receive beamforming vectors,
\begin{equation} \label{e6}
\boldsymbol{h}^{s,K^2} =
\begin{bmatrix}
    (\boldsymbol{w}^s_1)^H\boldsymbol{H}\boldsymbol{f}^s_1  \\
    (\boldsymbol{w}^s_2)^H\boldsymbol{H}\boldsymbol{f}^s_1   \\
    \vdots  \\
    (\boldsymbol{w}^s_1)^H\boldsymbol{H}\boldsymbol{f}^s_2  \\
    (\boldsymbol{w}^s_2)^H\boldsymbol{H}\boldsymbol{f}^s_2   \\
    \vdots  \\
    
    (\boldsymbol{w}^s_K)^H\boldsymbol{H}\boldsymbol{f}^s_{K}
\end{bmatrix}.
\end{equation}
In order to find the desired beamforming vectors, the dictionary matrix of all the possible steering vectors for the angles is defined as 
\begin{equation} \label{e7}
\boldsymbol{A}_{DIC} = [ \boldsymbol{a}(0), \boldsymbol{a}(\frac{2\pi}{N}),\dots, \boldsymbol{a}(\frac{2\pi(N-1)}{N})].
\end{equation}
Finding the beamforming vector for the $k$th sub-range at the TX is performed by computing
\begin{equation} \label{e8}
\boldsymbol{A}_{DIC}^H\boldsymbol{f}_k^{s} = \boldsymbol{z}_i^{s,k},
\end{equation}

 \begin{equation} \label{e9}
\boldsymbol{z}_i^{s,k}=\left\{\begin{array}{cl}
C_s,&\mbox{if }\frac{i2\pi}{N}\mbox{ is in the subrange}\\
0,& \mbox{elsewhere}\end{array}\right.
\end{equation}
where $C_s$ is a constant that is chosen to make the magnitude of the beamforming vectors equal to one ($\|\boldsymbol{f}\| = 1$ ). From equation (\ref{e8}), $\boldsymbol{f}_k^{s}$ is calculated as 

\begin{equation} \label{e10}
 \boldsymbol{f}_k^{s} = (\boldsymbol{A}_{DIC}\boldsymbol{A}_{DIC}^H)^{-1}\boldsymbol{A}_{DIC}\boldsymbol{z}^{s,k}.
\end{equation}
The same procedure is used to find the beamforming vectors of the RX. After $K^2$ measurement, the RX will compare the magnitude of $K^2$ received pilots and choose the one with the largest magnitude which is the desired path. In other words, we estimate the AoD ($\hat{k_t}$) and AoA ($\hat{k_r}$) based on the corresponding pilot.

\section{Feedback reduction algorithms ensuring PEE}

In this section, first, we introduce the sparse representation of the communication system. Next, we explain the maximum likelihood detection (MLD) method which is used in the algorithms for channel estimation. Then, the optimal number of feedback is explained, and finally, the RAF algorithm is presented.

\subsection{A sparse representation of the system}
A new matrix $\boldsymbol{G}^{M}$ at its initial state is defined as 

\begin{equation} \label{e11}
 \boldsymbol{G}^{M}=  \boldsymbol{G}^{K^2}  =  \mathbf{I}_{K \times K},
\end{equation}
The index $M$ used in the notation denotes the number of measurements. After initial channel estimation, $M$ is equal to $K^2$. Substituting equation (\ref{m6}) into (\ref{e6}), we have 
\begin{equation} \label{e12}
\boldsymbol{h}^{s,K^2} =
\begin{bmatrix}
    (\boldsymbol{w}^s_1)^H\boldsymbol{H}\boldsymbol{f}^s_1  \\
    (\boldsymbol{w}^s_2)^H\boldsymbol{H}\boldsymbol{f}^s_1   \\
    \vdots  \\
    (\boldsymbol{w}^s_1)^H\boldsymbol{H}\boldsymbol{f}^s_2  \\
    (\boldsymbol{w}^s_2)^H\boldsymbol{H}\boldsymbol{f}^s_2   \\
    \vdots  \\
    
    (\boldsymbol{w}^s_K)^H\boldsymbol{H}\boldsymbol{f}^s_{K}
\end{bmatrix}
= 
 N \alpha
\begin{bmatrix}
    (\boldsymbol{w}^s_1)^H\boldsymbol{a}_r(\theta) \boldsymbol{a}_t^H(\phi) \boldsymbol{f}^s_1  \\
    (\boldsymbol{w}^s_2)^H\boldsymbol{a}_r(\theta) \boldsymbol{a}_t^H(\phi) \boldsymbol{f}^s_1   \\
    \vdots  \\
    (\boldsymbol{w}^s_1)^H\boldsymbol{a}_r(\theta) \boldsymbol{a}_t^H(\phi) \boldsymbol{f}^s_2  \\
    (\boldsymbol{w}^s_2)^H\boldsymbol{a}_r(\theta) \boldsymbol{a}_t^H(\phi) \boldsymbol{f}^s_2   \\
    \vdots  \\
    
    (\boldsymbol{w}^s_K)^H\boldsymbol{a}_r(\theta) \boldsymbol{a}_t^H(\phi) \boldsymbol{f}^s_{K}
\end{bmatrix}
\end{equation}
The multiplication of $(\boldsymbol{w}^s_1)^H\boldsymbol{a}_r(\theta)$ and $\boldsymbol{a}_t^H(\phi) \boldsymbol{f}^s_1$ is only non-zero if the AoA and AoD are aligned to the beamforming vectors. Therefore, only one row of $\boldsymbol{h}^{s,K^2}$ is non-zero. Finding the AoA and AoD is equivalent to finding a $K^2 \times 1$ vector $\boldsymbol{v}$ which is zero everywhere except the desired row of $\boldsymbol{G}^{M}$ where it is equal to one. Hence, $\boldsymbol{h}^{s,K^2}$ and the observation vector can be written as 

\begin{equation} \label{e13}
 \boldsymbol{h}^{s,K^2} = xNC_s^2 \alpha  \boldsymbol{G}^{M}\boldsymbol{v}^T
\end{equation}
\begin{equation} \label{e14}
\boldsymbol{y}^{M} = xNC_s^2 \alpha  \boldsymbol{G}^{M}\boldsymbol{v}^T + \boldsymbol{n}^{M}.
\end{equation}
Assuming element $d$ of $\boldsymbol{v}$ is one, the estimated AoA ($\hat{k}_t$) and AoD ($\hat{k}_r$) are calculated by
\begin{equation} \label{angles}
\hat{k}_t = \lceil \dfrac{d}{K} \rceil ,\; \hat{k}_r = d - K(\hat{k}_t - 1).
\end{equation}
The new presentation of the system indicates that the possible outcomes of the channel estimation are equivalent to the rows of matrix $\boldsymbol{G}^{M}$.

\subsection{Maximum Likelihood Detection (MLD)}

In our algorithms, the MLD method will be used for the estimation of AoA and AoD. In this section, the method is described. After M measurement, the distribution of observation vector $\boldsymbol{y}^{M}$ can be written as 
\begin{equation} \label{e15}
    \boldsymbol{y}^{M} = \mathcal{C}\mathcal{N}(0, \boldsymbol{\Sigma_v} ),  
\end{equation}
where 
\begin{align} \label{e16}
\boldsymbol{\Sigma}_{v} &= PN^2C_s^4\boldsymbol{G}^{M}\boldsymbol{v}\boldsymbol{v}^T(\boldsymbol{G}^{M})^H + N_0 \boldsymbol{I}_{M}.
\end{align}
We refer to \cite{overlap} for the derivation. It can be seen that the received vector follows circularly symmetric complex Gaussian (CSCG) distribution which has the distribution of
\begin{align} \label{e17}
    f(\boldsymbol{y}^{M}&|\boldsymbol{v},\boldsymbol{G}^{M}) =\\ & \frac{1}{\pi^{K^2}\text{det}(\boldsymbol{\Sigma}_{v})} \text{exp}(-(\boldsymbol{y}^{M})^H \boldsymbol{\Sigma}^{-1}_{M} {\boldsymbol{y}^{M}}). \nonumber
\end{align}
In order to get a better understanding of the probability density function, it is useful to see them in terms of probability. Defining the set $\mathcal{V}$ as all possible $K^2$ outcomes of the vector $\boldsymbol{v}$, the probability can be written as 

\begin{align} \label{18}
   p(\boldsymbol{v}|\boldsymbol{y}^{M})  =  \frac{ f(\boldsymbol{y}^{M}|\boldsymbol{v}) }{ \sum\limits_{\boldsymbol{w}\in {\mathcal{V}} }  f(\boldsymbol{y}^{M}|\boldsymbol{w})   }. 
\end{align}

\subsection{Optimal number of feedback to achieve the PEE}

In order to have a benchmark to compare the RAF algorithm's feedback performance, we need to know what the optimal number of feedback bits is. Note that the number has to ensure the desired PEE. In other words, we are looking for the minimum implementable feedback bits number from the RX that guarantees the desired PEE. From information theory, we know that the minimum number is one with a single feedback including $\lceil\log_2(K)\rceil$ bits\cite{ref10}. We verify that this number is actually achievable by developing an algorithm which only needs $\lceil\log_2(K)\rceil$ bits feedback. The cost of having the optimal number of feedback bits is a large number of channel measurements. Therefore, this algorithm is just used as a benchmark and can not be a good alternative in practice.

\begin{figure}[t!]
\centering
\includegraphics[width=2.8in]{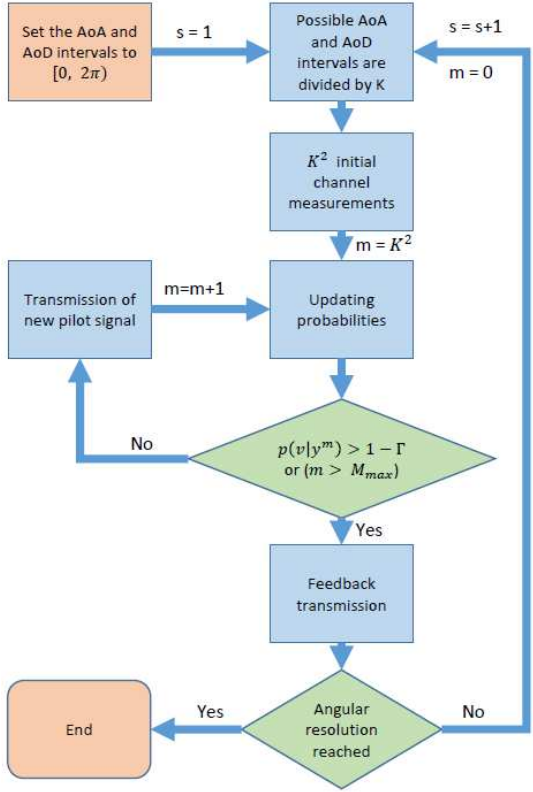}
\caption{Flow Diagram of an algorithm which requires the optimal number of feedback bits.}
\label{SF}
\end{figure} 

We denote $\Gamma$ as the probability of the event that a channel estimation is incorrect. The algorithm starts by having the initial $K^2$ measurements which result in the primary channel estimation. The TX continues to send the pilots using the same sequence as the initial measurements. After each transmission, using MLD the RX calculates $p(\boldsymbol{v}|\boldsymbol{y}^{M})$. As soon as reaching the desired PEE ($p(\boldsymbol{v}|\boldsymbol{y}^{M})> (1 - \Gamma)$) the RX will feedback $\lceil\log_2(K)\rceil$ bits to notify the TX about the estimated AoD. The process of adding a new measurement is written mathematically as 
\begin{align}\label{19}
\begin{split}
\boldsymbol{y}^{M+1} = \sqrt{P} x \left[\begin{array}{ccc} \boldsymbol{h}^{M} \\ (\boldsymbol{w}_{M/K}^{s})^H  \boldsymbol{H} \boldsymbol{f}_{\mod(M+1,K)+1}^{s} \end{array}
\right]\\ +\left[\begin{array}{ccc} \boldsymbol{n}^{(M)}  \\ (\boldsymbol{w}_{M/K}^{s})^H  n   \end{array}\right],
\end{split}
\end{align}
where $n$ is the noise on the $(M+1)$-th measurement. Note that there is always a probability of ‘outage' when the channel coefficient is close to zero. In order to prevent the excessive number of measurement, we set a maximum to the number of pilots which can be transmitted denoted by $M_{max}$. The complete flowchart of the algorithm is given in Fig. \ref{SF}.

\subsection{Robust adaptive multi-feedback algorithm (RAF) }
Multi-stage channel estimation algorithms are mainly based on a fixed number of channel estimation. As an example, authors in \cite{RH} used $K^2$ measurements in each stage to estimate the channel. Although the proposed algorithms are effective, they did not derive the performance in terms of the PEE. If due to the additive noise, the detection of the estimated AoA and AoD is incorrect in any of the stages, the algorithms will not be able to estimate the channel correctly. Therefore, devising an algorithm to ensure the PEE is crucial. Authors in\cite{RACE}, proposed a rate adaptive algorithm (RACE) in order to reach the desired PEE. Unfortunately, the algorithm requires a large number of feedback bits even for $K=2$, particularly in low SNR. 
We propose the RAF algorithm. In contrary to the existing algorithms, RAF exploits the estimated channel coefficient. As it will be illustrated, the significance of using channel coefficient is the entailed information about the number of measurements required for the channel estimation. This helps to estimate the time to commence sending the feedback bits from the receiver. The algorithm requires a low feedback overload and pilot transmissions.

Before explaining the algorithm we use information theory to find a lower bound for the number of measurements. The channel estimation is equivalent to finding a vector  $\boldsymbol{v}$ which contains $K^2$ binary bits encoded into $M$ (number of pilots transmitted) symbols. Therefore, the system has a transmission rate of $ \mathcal{C} = \dfrac{K^2}{M}$. According to the Shannon-Hartley theorem\cite{ref10} 

\begin{align} \label{e20}
   \mathcal{C}  &= \dfrac{K^2}{M} \leq \log_2(1 + SNR)\\
   \label{e23}
   &\rightarrow M \geq \dfrac{K^2}{log_2(1 + SNR)},
\end{align}
where $SNR_s$ (in stage s) can be written as
\begin{align} \label{e22}
	SNR_s = \dfrac{|\alpha|^2PK^{(2s-2)}}{N_0}.
\end{align}
Substituting equation (\ref{e22}) in (\ref{e23}), a lower bound can be found for the number of measurements that are required in each stage on condition of the estimated value of $\alpha$. After $M$ measurements ($M \geq K^2$), if the mean of observations received in the estimated AoA and AoD are denoted by $Y^M$, the value of $\alpha$ can be estimated as 

\begin{align}\label{e21}
\hat{\alpha} =  \frac{ Y^M}{\sqrt{P}N C^2_{s} }.
\end{align}
Therefore, we have a lower bound for the number of measurements required.

\begin{figure}[t!]
\centering
\includegraphics[width=3.4in]{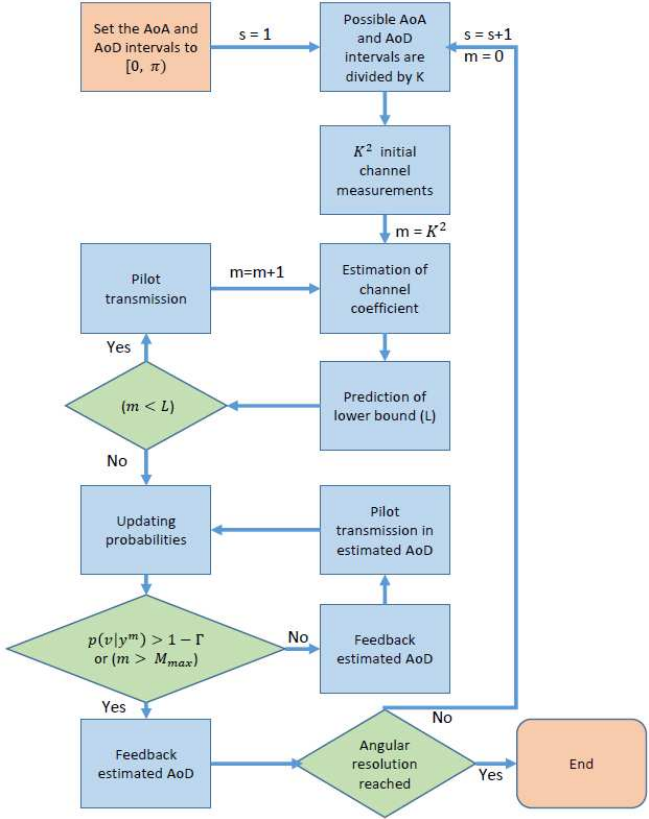}
\caption{Flow Diagram of the Proposed RAF Algorithm.}
\label{Shannon}
\end{figure}

In each stage, the algorithm starts by conducting $K^2$ initial channel measurements. The MLD enables the system to have an estimation of the AoA and AoD which can be used to estimate the value of channel coefficient ($\alpha$). Having the estimated $\alpha$, the receiver can predict a lower bound for the required number of measurements. Up to the point of reaching the PEE threshold, the TX continues to send the pilots as explained in the optimal feedback algorithm. As the pilots are accumulated, the same process of MLD is used to achieve a better estimation of $\alpha$ which results in obtaining a more accurate lower bound. After reaching the PEE threshold, the RX feeds back the estimated AoD. At this point, the TX stops sending the pilots in the order of initial channel estimation and only sends a pilot in the estimated AoD. The RX knows the estimated AoA and utilizes the corresponding combiner to receive the pilot. Following the same process after receiving each pilot, the RX estimates the AoA and AoD and feeds back the estimated AoD. This procedure is terminated as soon as the required estimation precision is reached. In the final transmission of feedback, an extra bit will be transmitted to notify the transmitter to stop the transmission of pilot signals. The flowchart of RAF algorithm is shown in Fig. \ref{Shannon}.

%
%

\section{simulation results and discussions}
In this section, we show numerical results to demonstrate the effectiveness of the proposed algorithm. The system is assumed to have $N=64$ antennas at both the TX and RX. The channel coefficient is assumed to follow a Gaussian distribution with zero mean and unit covariance ($\mathcal{C}\mathcal{N}(0,1)$). Since the algorithm follows the same process in each stage, we compare the results of the different algorithms in a single stage. The results can be scaled for all of the stages. In order to compare the results to \cite{RH} and \cite{RACE}, we use $K=2$. The target PEE is set to $ 10^{-2}$. Finally, the maximum number of measurements is set to $M_{max} = 264$ for all the algorithms.

\subsection{ PEE performance and its comparison to prior work}
\begin{figure}[t!]
\centering
\includegraphics[width=3.548in]{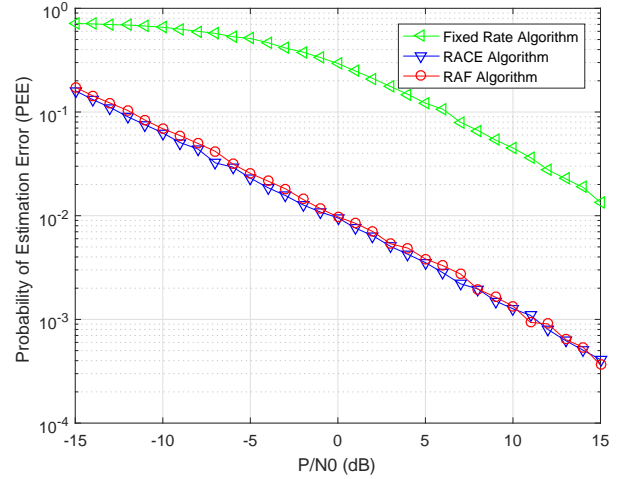}
\caption{Performance of the RAF algorithm compared to the algorithms in \cite{RACE} and \cite{RH} in terms of PEE.}
\label{pee}
\end{figure}

Fig. \ref{pee} represents the probability of estimation error for different signal to noise ratios. The results are compared to two prior work \cite{RH} and \cite{RACE}. In \cite{RH}, the authors were only concerned with the channel estimation. As it can be seen from Fig. \ref{pee}, the algorithm can not ensure that the probability of estimation error stays below the threshold. This algorithm is used as a baseline in this figure. The main comparison is between \cite{RACE} and our work. The figure indicates that both algorithms achieve the desired PEE. Note that there is always a probability of outage in the system. That is the reason why in low SNR the PEE is over the predetermined threshold. The difference between the achieved PEE of the two algorithms is negligible. 

\subsection{Overall channel estimation time and its comparison to prior work}

Each pilot transmission requires one time slot. On the other hand, each feedback also consists of one bit of information transmitted in one time slot. Therefore, the overall time of the channel estimation is the sum of these two numbers. In order to evaluate the overall performance, these two factors need to be studied. Recall that the algorithms are ensuring the PEE in addition to channel estimation. Here, we see how the RAF algorithm significantly decreases the overall channel estimation time.
   
\begin{figure}[t!]
\centering
\includegraphics[width=3.548in]{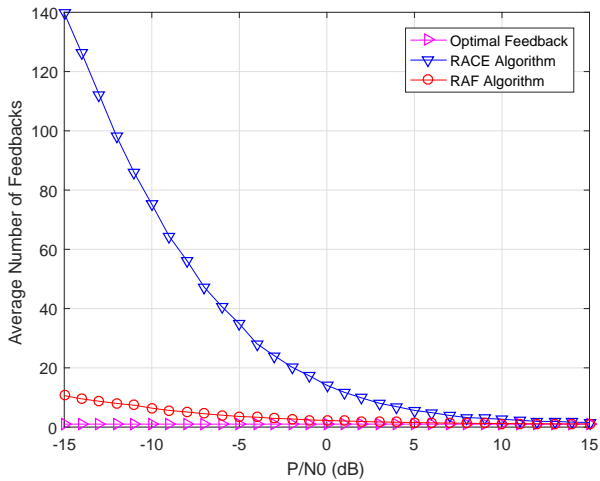}
\caption{Feedback performance of the RAF algorithm compared to \cite{RACE} and the optimal number of feedbacks.}
\label{feedback}
\end{figure}

\begin{figure}[t!]
\centering
\includegraphics[width=3.548in]{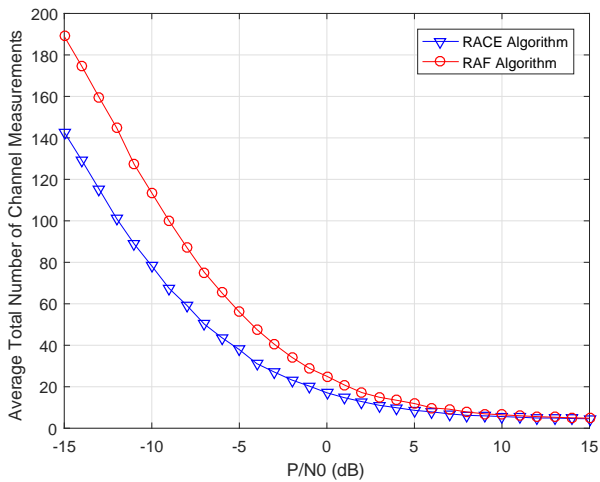}
\caption{Performance of the RAF algorithm compared to the algorithm in \cite{RACE} in terms of the average number of measurements required in each stage.}
\label{nom}
\end{figure}

We compare our results in terms of the feedback overhead and number of pilot transmissions (measurements) with prior work in \cite{RACE}. Fig. \ref{feedback} explains the performance of the algorithms for the number of feedback bits they need. It can be seen that the RAF algorithm requires a significantly low number of feedback. The average feedback bits required is almost as low as the optimal number. The difference between the algorithms becomes obvious particularly in low SNR regime. In terms of the number of pilot transmission, the RAF algorithm demonstrates a slightly higher performance (Fig. \ref{nom}).
 
\begin{figure}[t!]
\centering
\includegraphics[width=3.548in]{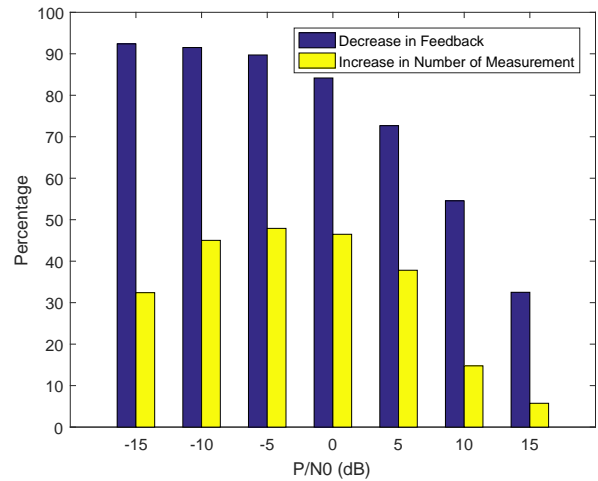}
\caption{Change in the performance of the RAF algorithm in comparison with \cite{RACE}. The graph explains the trade-off between the rise in the average number of measurements and feedback reduction.}
\label{to}
\end{figure}

\begin{figure}[t!]
\centering
\includegraphics[width=3.548in]{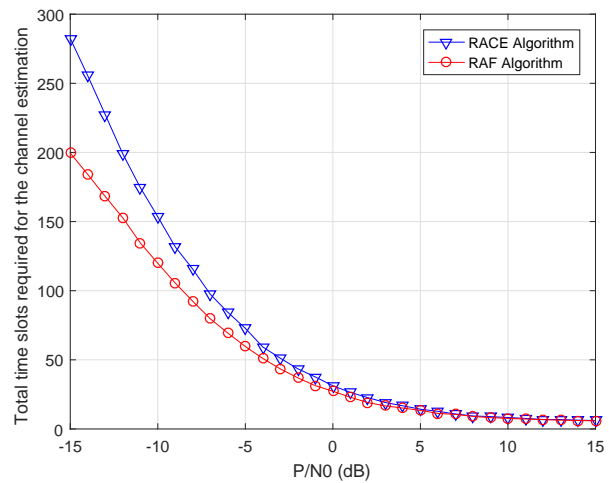}
\caption{Comparison of the time required for the channel estimation in each stage.}
\label{s}
\end{figure}

The trade-off between the number of measurements and feedback overhead is explained in Fig. \ref{to}. It is clear to see that the reduction in feedback bits is much higher than the increase of pilot transmissions required for the channel estimation. On average, the feedback reduction is $75.5\%$, whereas the increase in the number of measurement is only $34.4\%$.  The overall time of the channel estimation in each stage of the algorithms is shown in Fig. \ref{s}. This figure illustrates the superior performance of the RAF algorithm. On average, from SNR of -15dB to 15dB, the performance improves by $14\%$. At low SNRs, the difference is more significant. For instance, in an SNR of -15dB, the overall time required for the channel estimation is $30\%$ reduced using the RAF algorithm.

\section{conclusion }
In this paper, a robust adaptive multi-stage algorithm called 'RAF' was proposed to reduce the high overhead of channel estimation feedback in the existing algorithms. The RAF algorithm exploits the estimation of channel coefficient to find a lower bound for the number of measurements. Hence, it is able to predict when to start the feedback transmission which results in a significant reduction of feedback overhead. The simulation results indicate that the algorithm reduces the time required for the channel estimation On average by $14\%$ and the feedback overhead by $75.5\%$.

\bibliographystyle{IEEEtran}
\bibliography{CS_paper}


\end{document}